\documentclass[onecolumn,useAMS]{mn2e}
\usepackage{graphicx}
\usepackage{amsmath, amssymb}

\begin{document}
\title[Self-consistent vertical response of a galactic disc]
{Self-consistent response of a galactic disc to vertical perturbations}
\author[K. Saha and C.J. Jog]
       {Kanak Saha\thanks{E-mail : kanak@physics.iisc.ernet.in, 
       cjjog@physics.iisc.ernet.in} and
	Chanda J. Jog \\
   Department of Physics,
Indian Institute of Science, Bangalore 560012, India \\}

\maketitle

\begin{abstract} We study the self-consistent, linear response of a galactic 
disc to vertical perturbations, as induced say by a tidal interaction. We
calculate the self-gravitational potential corresponding to a non-axisymmetric, self-consistent density response of the disc using the Green's function method. The response potential is shown to oppose the perturbation potential because the self-gravity of the disc resists the imposed potential,  and this resistence is stronger in the inner parts of a galactic disc.  For the $m=1$ azimuthal
 wavenumber, the disc response opposes the imposed perturbation upto 
a radius that spans a range of 4-6 disc scale-lengths, so that the disc shows a net warp only beyond this region.
This physically explains the well-known but so far unexplained observation 
(Briggs 1990) that warps typically set in beyond this range of radii. We show that the
inclusion of a dark matter halo in the calculation only marginally changes (by $\sim 10\%$) the radius for the onset of warps. For perturbations with higher azimuthal wavenumbers, the net signature of the vertical perturbations can only be seen at larger radii - for example beyond 7 exponential disc scale-lengths for $m=10$. Also, for high $m$ cases, the magnitude of the negative disc response due to the 
disc self-gravity is much smaller. This is shown to result in corrugations of the mid-plane density, which explains the puzzling scalloping with $m=10$ detected in HI 
in the outermost regions $\sim 30 kpc$ in the Galaxy by Kulkarni et al. (1982). 
\end{abstract}


\section{Introduction}

Galaxy interactions including interactions with satellite
galaxies are now known to be common. As a consequence of such a tidal 
interaction, vertical distortions in the galactic discs can be easily set in.
The most common kind of vertical distortions in discs are warps,
corresponding to the azimuthal wavenumber $m=1$ .
Tidal interactions with neighbouring galaxies
have often been suggested as the mechanism for the origin of
warps (Schwarz 1985, Zaritsky \& Rix 1997). The generation of the warp of 
our Galaxy due to the tidal interaction with the 
Large Magellanic Clouds(LMC) was studied by Hunter \& Toomre (1969), and more 
recently by Weinberg (1995) in the presence of an active dark matter halo.
Recent statistical studies by Sanchez-Saavedra et al. (1990) and 
Reshetnikov \& Combes (1998) conclude that at least $50\%$ of all the 
observed galaxies show warping in their discs. Their study on environmental 
effects, along with another recent analysis of a large number of edge-on 
spiral galaxies in the R-band (Schwarzkopf \& Dettmar 2001), suggests that 
tidal perturbation (the $m=1$ component) plays an important role 
in generating and influencing large-scale warps.

 While it is clear that a warp is more likely to set in in the
outer parts of a galaxy where the tidal force is stronger, it is 
 not yet known  what decides the actual radius at which a warp starts 
 in the disc.  It is a well-known observational fact that warps are seen in
the outer regions of galaxies, typically starting at $\sim 4-6$
exponential disc scale-lengths, as was first seen in HI observations by 
Briggs (1990). However there exists no clear physical explanation for the  
radius at which a warp sets in. In fact, little theoretical attention has been devoted to this question.

Other kinds of vertical distortions that are seen are small-scale 
corrugations in HI in our Galaxy (Quiroga 1974)  or in stars in
external galaxies (Florido et al. 1991),
and also scalloping in HI in the outer Galaxy
 (Kulkarni et. al. 1982). Again, the  physical origin of these
non-axisymmetric features is not understood.

In this paper, we address these questions and show that the
self-gravity of the disc resists any vertical distortion, especially in
the inner regions. Thus using very general physical principles, we 
derive semi-analytically a minimum radius for the onset of each of the above
vertical perturbations of a galactic disc.
Each Fourier component of a perturbing potential will generate its own 
signature in the galactic disc. Observations of these signatures in the disc 
depend on the strength of the perturbation component and the disc response 
corresponding to it.
The behaviour of a disc subjected to an external perturbation can 
entirely be studied by its response funtion. 

We study here the response of an axisymmetric disc, perturbed by a 
tidal potential and show that the disc self-gravity plays a significant 
role in sketching a finger-print of the Fourier component of the perturbation 
onto the disc. We obtain the self-gravitational potential corresponding to a 
non-axisymmetric, self-consistent, density response of the disc induced by 
the perturbation component and show that it opposes the perturbing potential. 
The resulting negative disc response (see Jog 1999 for the planar case) 
decreases the strength of the perturbation in the disc. The vertical linear 
disc response obtained here is shown to oppose the perturbation
upto a certain radius and for the $m=1$ component, in particular, 
it is upto about $4$ disc scale-lengths. 
Thus the net disc response for $m=1$ or warps can only be seen beyond 
this radius, which is in a fairly good agreement with observations. 
We also add a rigid dark matter halo and show that it only marginally 
affects the location of the onset of warps.

For perturbations with higher azimuthal wavenumbers ($m$) the disc is 
shown to be well resistant upto higher radii. In particular 
we show that any signature of the $m=10$ component of the 
perturbation can only be observed in the very outer region of the disc 
beyond $\sim 7$ disc scale-length for 
an exponential disc.

In Section 2, we describe the treatment for the self-consistent linear response 
of the disc. The results and the comparison with observations
are given in Section 3, and the conclusions are summarised in Section 4.

\bigskip

\section{Vertical dynamics} 

We study the vertical response in a disc galaxy subjected to a general perturbing 
potential of $\cos (m\varphi)$ type. We use the cylindrical 
coordinate ($R,\varphi,z$) 
geometry which is the most natural for describing dynamics in a disc. 
The dynamical 
system under study thus consists of a disc embedded in a dark matter halo.

The density distribution of the unperturbed axisymmetric disc is given by

$$ \rho_{d}(R,z) \: = \: \rho_{d0}e^{-R/R_d}e^{-{|z|}/z_{\circ}} \eqno(1) $$ 

\noindent where $\rho_{d0}$ is the disc central density and $R_d$ is the disc 
scale-length. $z_{\circ}$ is the vertical scale-height which in general is a 
function of the galactocentric distance $R$. In an interacting or merging 
disc the scale-height $z_{\circ}$ increases systematically with the radial 
distance. The choice of the double exponential disc comes from a recent 
study by Schwarzkopf \& Dettmar (2001) of tidally-triggered vertical disc perturbations performed on a fairly large sample of disc galaxies which includes both interacting as well as non-interacting galaxies. 
Their study reports that $33\%$ of the discs show an exponential 
mass distribution in the vertical direction indicating that most of the 
disc galaxies prefer to have such non-isothermal vertical profiles. 
So we are interested in investgating how these non-isothermal 
discs respond to an external perturbation of the type mentioned above.

The dark matter halo is modeled as an axisymmetric spheroidal system with a pseudo-isothermal density profile which gives an asymptotically flat rotation curve( de Zeeuw \& Pfenniger 1988 ):

$$ \rho_{h}(R,z) \:=\: \frac{\rho_{h0}}{1 + (R^2 + \frac{z^2}{q^2})/R_{c}^2} \eqno(2)$$ 
\noindent where $\rho_{h0}$ is the central density of the dark matter halo, $R_c$ is the core radius and $q$ is the halo flattening parameter.

The stars in the disc midplane ($z=0$) move in a circular orbit at 
$R=R_{\circ}$ with an angular velocity $\dot{\varphi_{\circ}}=\Omega_{\circ}$. So the unperturbed circular orbit in the $z=0$ plane is defined by $R=R_{\circ}$ and $\varphi_{\circ}=\Omega_{\circ}t$. The angular velocity $\Omega_{\circ}$ is given by

$${\Omega_{\circ}}^2 \:=\: {\frac{1}{R_{\circ}}} \left.{\frac{\partial{\Psi_{\circ}}}{\partial{R}}}\right|_{R=R_{\circ}} \eqno(3a)$$

\noindent where $\Psi_{\circ}$ is the total unperturbed potential:

$$\Psi_{\circ}=\Psi_{d\circ} + \Psi_{h\circ} \eqno(3b)$$
\noindent $\Psi_{d\circ}$ and $\Psi_{h\circ}$ are connected to the mass distribution described by eq.[1] and eq.[2] respectively through the Poisson equation. 
 $\Psi_{d0}$ is given by (see Sackett \& Sparke 1990, Kuijken \& Gilmore 1989):
$$ \Psi_{d\circ}(R,z) \:=\: -4\pi G{\rho_{d0}}{R_{d}}^2 z_{\circ} {\int_{0}^{\infty}}{\frac{J_{0}(k R)}{[1+(k R_{d})^2]^{3/2}}} {\frac{ {{k z_{\circ}}e^{-|z|/z_{\circ}}} - {e^{-k |z|}} } {k^2 {z_{\circ}}^2 -1}}dk \eqno(3c)$$

A similar calculation for the halo is more cumbersome and we have used the formula given by eq.(2.88) in Binney \& Tremaine (1987) to calculate the gravitational force generated by the spheroidal dark matter halo described by the screened isothermal density distribution (eq.[2]).  

\noindent We consider the perturbing potential at any point($R,\varphi,z$) in the disc due to a perturber at a distance $D$ from the disc centre and at an inclination angle $i$ with respect to the disc midplane:
$$ \Psi_{1}(R,\varphi,z,\Omega_{p}) \:=\: \Psi_{p\circ}(R)(1+ {\frac{z}{D}}\sin{i}) \cos(m\varphi -m\Omega_{p}t) \eqno(4a)$$

The above form of the potential is derived by assuming that $R/D<<1$ and $z/D<<1$ . 
\noindent The form of $\Psi_{p\circ}(R) \sim G M_p R/D^2$ can be taken as a constant in the above limit, where $M_p$ is the mass of the perturber. $\Omega_{p}$ is the pattern frequency of the perturbing potential. 

\noindent In the present study we consider only the disc response under the perturbing potential with zero-forcing frequency. Generally the zero-frequency response of the disc is much more pronounced than the response generated by a non-zero finite-frequency perturbation ( see Terquem 1998 in case of an accretion disc response ).
Thus the perturbing potential with zero-forcing frequency reduces to 
$$\Psi_{1}(R,\varphi,z) \:=\: \Psi_{p\circ}(R)(1+{\frac{z}{D}}\sin{i}) \cos(m\varphi) \eqno(4b) $$  
Note that potential of this form in case of $m=1$ generates a well-defined integral-sign warp in the disc.  

\noindent The total potential that a disc particle experiences is given by
$$\Psi_{total}(R,\varphi,z) \:=\: \Psi_{\circ}(R,z) + \Psi_{1}(R,\varphi,z) \eqno(4c)$$

\subsection{Disc linear response:}

\subsubsection{Basic equations:}

The vertical response of a disc to an imposed external perturbation is governed by the following equations.
\noindent The vertical equation of motion in the small-slope approximation (i.e. for a small deviation from the disc midplane $z=0$) is given by

$$\frac{d^2{z_{m}}}{dt^2} \:=\: -{\nu_{\circ}}^2 z_{m} + \left.{\frac{\partial{\Psi_{1}}}{\partial z}}\right|_{({R_{\circ}},\varphi_{\circ},0)} \eqno(5a) $$
where the vertical frequency due to the unperturbed disc plus halo system is given by
$$ {\nu_{\circ}}^2 \:=\: \left.{\frac{\partial^2{\Psi_{\circ}}}{\partial{z^2}}}\right|_{(R=R_{\circ},0)} \eqno(5b) $$

\noindent 
The linearized equation of continuity, connecting the perturbed velocity to the perturbed density, is given by ( Binney \& Tremaine 1987 )

$$\frac{\partial{\rho_{1}}}{\partial{t}} + \vec{\nabla}.({\rho_{\circ}}\vec{v_{1}}) + \vec{\nabla}.({\rho_{1}}\vec{v_{\circ}}) \:=\: 0 \eqno(6a)$$

\noindent We denote $v_{1}=v_{z m}=dz_{m}/dt$ as the perturbed z-velocity and $v_{\circ}=v_{z0}$ as the unperturbed z-velocity. For simplicity we assume $v_{z0}=0$.

\noindent Then the continuity equation (6a) reduces to
$$\frac{\partial{\rho_{1}}}{\partial{t}} + \vec{\nabla}.({\rho_{\circ}}\vec{v}_{zm}) \:=\: 0 \eqno(6b) $$

\noindent The response potential corresponding to this perturbed density is connected through the Poisson equation as
$$ \nabla^2 {\Psi_{resp}}(R,\varphi,z) \:=\: 4\pi G \rho_{1}(R,\varphi,z) \eqno(7) $$

\noindent 
We solve the above three equations namely eqs.(5a), (6b) and (7) in order to obtain the disc response potential.

On substituting the perturbing potential eq.[4b] into eq.[5a] we obtain a forced oscillator-type second-order differential equation and the solution can be written as

$$z_{m} \:=\: H(R_{\circ})\cos(m \Omega_{\circ} t) \eqno(8) $$
\noindent Where $$ H(R_{\circ}) \:=\: \frac{\left.{\frac{\partial{\Psi_{1}}}{\partial z}}\right|_{({R_{\circ}},0)}} {{\nu_{\circ}}^2 - {\Omega_{\circ}}^2} \eqno(9) $$

\noindent So the perturbed velocity in the z-direction is given by
$$ \vec{v}_{z m} \:=\: -m{\Omega_{\circ}}H(R_{\circ}) \sin(m \Omega_{\circ} t) \hat{z} \eqno(10) $$

\noindent Substituting this into eq.[6b] and solving for $\rho_{1}$ we obtain
$$\rho_{1} \:=\: -{\frac{\partial {\rho_{\circ}}}{\partial z}} H(R_{\circ})\cos(m \Omega_{\circ} t) $$

\noindent Or
$$\rho_{1}(R_{\circ},\varphi_{\circ},z) \:=\: -{\frac{\partial {\rho_{\circ}}}{\partial z}} H(R_{\circ})\cos(m \varphi_{\circ}) \eqno(11) $$

\noindent From now on we will use the general co-ordinates $(R,\varphi,z)$ 
instead of $(R_{\circ},{\varphi_{\circ}},z)$. Note that the unperturbed 
density distribution for the disc plus the halo system is given by 
${\rho_{\circ}}(R,z) ={\rho_{d}}(R,z) + {\rho_{h}}(R,z)$ (from eqs.[1] and 
[2]). So that after subsituting $\rho_{\circ}$ in eq.[11] we obtain

$$\rho_{1}(R,\varphi,z) \:=\: [{\rho_{d}}(R,z) + \zeta_{h}(R,z)]\left\{\frac{H(R)}{z_{\circ}}\right\} \cos(m \varphi) \eqno(12) $$ 

\noindent In the above equation $\zeta_{h}(R,z) = -z_{\circ}\partial 
{\rho_{h}}/{\partial z}$ which is a positive quantity for any centrally 
concentrated self-gravitating system. 
Since ${\partial{\Psi_{1}}}/{\partial z} > 0$ 
and $\nu_{\circ}^2 > \Omega_{\circ}^2$ in an oblate geometry, the function 
$H(R) > 0$.  {\it Thus the perturbed response density follows the 
perturbing potential}. This behaviour is generally true for any 
self-gravitating, centrally 
concentrated disc (see Section 2.1.2 for details).
The total volume density in the disc due to the presence of the $m^{th}$ component of the perturbation can be written as

$$\rho_{m}(R,\varphi,z) \:=\:{\rho_{\circ}}(R,z)\left[1 +(\frac{\rho_{d}+\zeta_{h}}{\rho_{\circ}}){\frac{H(R)}{z_{\circ}}}\cos(m \varphi)\right] \eqno(13) $$

\subsubsection{Response potential:}
The solution of Poisson equation is one of the most tricky issues in this work. Now, the Poisson equation can be solved in two different ways: either by solving the partial differential equation which needs proper boundary conditions to 
 be imposed, or by using the integral equation i.e. using the Green's function method (Jackson 1975). For the planar case, an infinitesimally-thin disc approximation can be used and hence the differential approach
could be used (Jog 1999). This approximation is not valid here, and hence this approach cannot be used in the present
work. Instead here we have to solve the integral 
 equation using the Green's function method, for which the boundary conditions are directly in-built 
 into the integral equation (Arfken 1985).
  A compact form of the Green's funtion in cylindrical coordinates can be 
  found in Cohl \& Tohline (1999). We now solve the 
Poisson equation (eq.[7]) for the non-axisymmetric perturbed density given 
by eqn(12b) using the Green's function technique. So that

$$ \Psi_{resp}(R,\varphi,z) \:=\: -G {\int_{V}}\frac{\rho_{1}(\vec{r^{\prime}})}{|\vec{r} - \vec{r^{\prime}}|}{d^3}{r^{\prime}}  \eqno(14) $$

\noindent where
$${|\vec{r} - \vec{
r^{\prime}}|} \:=\:{\left[ R^2 + {R^{\prime}}^2 -2 R R^{\prime} \cos(\varphi^{\prime} - \varphi) + (z^{\prime} - z)^2 \right]}^{1/2} $$
 
\noindent Then
$$\Psi_{resp}(R,\varphi,z) \:=\: -G {\int_{V}}\frac{ \rho_{r}(R^{\prime},z^{\prime})\cos(m \varphi^{\prime}) R^{\prime} dR^{\prime} d{\varphi^{\prime}} d z^{\prime} }{ {\left[ R^2 + {R^{\prime}}^2 -2 R R^{\prime} \cos(\varphi^{\prime} - \varphi) + (z^{\prime} - z)^2 \right]}^{1/2} } \eqno(15) $$ 

\noindent In writing the above equation we have used
$\rho_{r}(R,z)=[\rho_{d}(R,z) + \zeta_{h}(R,z)] H(R)/z_{\circ}$.   

\noindent After performing the integration over $\varphi^{\prime}$ we obtain the response potential corresponding to an odd-m (i.e. m=1,3,5,7....) perturbation as

$${\Psi^{odd}_{resp}}(R,\varphi,z,m) \:=\: -4 G \cos(m \varphi) { {\int_{0}^{\infty}} {\int_{-\infty}^{\infty}} } { \frac{ \rho_{r}(R^{\prime},z^{\prime})R^{\prime} dR^{\prime} d z^{\prime} } {\sqrt{(R + R^{\prime})^2 + (z^{\prime} - z)^2 }} }\left[ 2 {\mathcal E}_{od}(m,k) - K(k)\right] \eqno(16) $$ 

\noindent where 
$${\mathcal E}_{od}(m,k) \:=\: {\int_{0}^{\pi/2}}\frac{{\sin^2 (m \beta)}}{\sqrt{1-k^2 {\sin^2(\beta)}}} d\beta \eqno(17a) $$
and K(k) is the complete elliptical integral of first kind and k
is as defined below
$$ k^2 \:=\: \frac{4 R R^{\prime}}{(R + R^{\prime})^2 + (z^{\prime} - z)^2 + {z_{s}}^2 } \eqno(17b) $$

\noindent In the above equation,  $z_{s}$ is the small softening parameter added to make the response potential regular at ${r}={r^{\prime}}$.

\noindent Similarly for even-m (i.e.,$ m=2,4,6,8$....) perturbation the response potential takes the form
$${\Psi^{even}_{resp}}(R,\varphi,z,m) \:=\: 
-4 G \cos(m \varphi) { {\int_{0}^{\infty}} {\int_{-\infty}^{\infty}} } 
{ \frac{ \rho_{r}(R^{\prime},z^{\prime})R^{\prime} 
dR^{\prime} d z^{\prime} } {\sqrt{(R + R^{\prime})^2 + (z^{\prime} - z)^2 }} }\left[ 2 {\mathcal E}_{ev}(m,k) - K(k)\right] \eqno(18) $$

\noindent where
$${\mathcal E}_{ev}(m,k) \:=\: {\int_{0}^{\pi/2}}\frac{{\cos^2 (m \beta)}}{\sqrt{1-k^2 {\sin^2(\beta) }}} d\beta \eqno(19) $$

\noindent Now substituting $\rho_{r}(R,z)$ into eq.(16) and eq.(18) we obtain

$${\Psi^{odd}_{resp}}(R,\varphi,z,m) \:=\: -4 G\cos(m \varphi) { {\int_{0}^{\infty}}}{\frac{H(R^{\prime})}{z_{\circ}}} {\mathcal Z}^{od}_{exp}(R,R^{\prime},z,m) R^{\prime} dR^{\prime} \eqno(20a) $$
\noindent where 
$${\mathcal Z}^{od}_{exp}(R,R^{\prime},z,m) \:=\: {\int_{-\infty}^{\infty}} { \frac{[\rho_{d}(R^{\prime},z^{\prime}) + \zeta_{h}(R^{\prime},z^{\prime})]}{\sqrt{(R + R^{\prime})^2 + (z^{\prime} - z)^2 }}}\left[ 2 {\mathcal E}_{od}(m,k) - K(k)\right] d z^{\prime} \eqno(20b) $$

\noindent and
$${\Psi^{even}_{resp}}(R,\varphi,z,m) \:=\: -4 G\cos(m \varphi) { {\int_{0}^{\infty}}}{\frac{H(R^{\prime})}{z_{\circ}}} {\mathcal Z}^{ev}_{exp}(R,R^{\prime},z,m) R^{\prime} dR^{\prime} \eqno(21a) $$

\noindent where 
$${\mathcal Z}^{ev}_{exp}(R,R^{\prime},z,m) \:=\: {\int_{-\infty}^{\infty}} { \frac{[\rho_{d}(R^{\prime},z^{\prime}) + \zeta_{h}(R^{\prime},z^{\prime})]}{\sqrt{(R + R^{\prime})^2 + (z^{\prime} - z)^2 }}}\left[ 2 {\mathcal E}_{ev}(m,k) - K(k)\right] d z^{\prime} \eqno(21b) $$

\noindent Using eq.(9) and the definition of the perturbing potential we have

$$ H(R^{\prime})\:=\: {\Psi_{p\circ}(R^{\prime})} h(R^{\prime}) \eqno(22a) $$

\noindent where $$ h(R^{\prime}) \:=\: \frac{\sin{i}}{\big[{\nu_{\circ}^{2}(R^{\prime}) - \Omega_{\circ}^2(R^{\prime})}\big]D} \eqno(22b) $$

\noindent We assume that $\Psi_{p\circ}(R^{\prime})$ is a very slowly varying function or even a constant (in agreement with the assumptions in sec 2.) over the disc size and has a small magnitude compared to the background, this is justified since the perturber is typically much more distant than the size of the disc. Thus we can take
this term  out of the integration in eqs.(20a) and (21a). Then the response potential in the disc midplane($z=0$) is given by

$${\Psi^{odd}_{resp}}(R,\varphi,z=0,m) \:=\: -4 G \rho_{d0}\cos(m \varphi) {{\Psi_{p\circ}(R)}} { {\int_{0}^{\infty}}}{ e^{-{R^{\prime}}/R_d}} {\frac{h(R^{\prime})}{z_{\circ}}} {\mathcal Z}^{od}_{exp}(R,R^{\prime},0,m) R^{\prime} dR^{\prime} \eqno(23a) $$ 

\noindent and
$${\Psi^{even}_{resp}}(R,\varphi,z=0,m) \:=\: -4 G \rho_{d0}\cos(m \varphi) {{\Psi_{p\circ}(R)}} { {\int_{0}^{\infty}}}{ e^{-{R^{\prime}}/R_d}} {\frac{h(R^{\prime})}{z_{\circ}}} {\mathcal Z}^{ev}_{exp}(R,R^{\prime},0,m) R^{\prime} dR^{\prime} \eqno(23b) $$
Note that the only significant contributions from the dark matter halo to the response potential come through the function $h(R)$ since at the $z=0$ plane $\zeta_{h}(R^{\prime},z^{\prime})$ part being an odd function of $z^{\prime}$ contributes zero to the $\mathcal{Z}$-integrals.

\noindent Substituting the perturbing potential at the $z=0$ plane into eqs. (23a) and (23b), we can write  the response potential for the odd-m and even-m perturbations as:

$${\Psi^{odd}_{resp}}(R,\varphi,z=0,m) \:=\: {\mathcal R}^{odd}_{m}(R){\Psi_{1}(R,\varphi,z=0)} \eqno(24) $$  
  
\noindent And

$${\Psi^{even}_{resp}}(R,\varphi,z=0,m) \:=\: {\mathcal R}^{even}_{m}(R){\Psi_{1}(R,\varphi,z=0)} \eqno(25) $$ 

\noindent where

$${\mathcal R}^{odd}_{m}(R) \:=\: -{4 G \rho_{d0}}{ {\int_{0}^{\infty}}} { e^{-{R^{\prime}}/R_d}} {\frac{h(R^{\prime})}{z_{\circ}}} {\mathcal Z}^{od}_{exp}(R,R^{\prime},0,m) R^{\prime} dR^{\prime} \eqno(26) $$

$${\mathcal R}^{even}_{m}(R) \:=\: -{4 G \rho_{d0}} { {\int_{0}^{\infty}}}{ e^{-{R^{\prime}}/R_d}} {\frac{h(R^{\prime})}{z_{\circ}}} {\mathcal Z}^{ev}_{exp}(R,R^{\prime},0,m) R^{\prime} dR^{\prime} \eqno(27) $$
  
\noindent  The above eqs.(26) and (27) define the dimensionless response potential of the disc for odd $m $ and for even $m$ perturbations respectively.

\noindent The numerical values of the functions 
${\mathcal Z}_{exp}(R,R^{\prime},0,m)$ are positive for both odd and even 
$m$ perturbations since at the midplane the dominant contribution 
from the function $\left[ 2 {\mathcal E}(m,k) - K(k)\right]$ ( for both even and odd cases ) comes when $R=R^{\prime}$ and 
$z^{\prime}=0$ i.e. $k \sim 1$. The function $h(R)$ is positive in an 
oblate dark matter halo (since $\nu_{\circ}^2 > \Omega_{\circ}^2$), whereas 
 it is negative when the dark matter halo is prolate 
(since $\nu_{\circ}^2 < \Omega_{\circ}^2$)
 and the disc self-gravity is negligible. 
Hence the response function in eqs.(26) and
(27) has a negative definite sign for the disc-alone system and also 
for a system consisting of a disc and an oblate dark matter halo. 
In contrast, this can have a positive sign for the case
of a disc plus a massive prolate dark matter halo. 
The studies of galactic structure and dynamics in the literature normally assume
 a spherical halo or an oblate dark matter halo, hence we have performed all our calculations for a nearly spherical halo i.e. with a very small obtaleness as a convenient choice. 

{\it Thus the dimensionless response potential has a sign opposite to the 
perturbation potential (eq.[4b]) in the commonly used two cases of
disc alone or a disc plus an oblate halo systems}. 
In the linear regime studied here the magnitude of the response potential 
is proportional to the strength of the perturbation potential. So 
the resulting disc response potential becomes weaker when the perturber is 
far away from the disc. 

\noindent A  similar negative disc response in the planar case for a
self-gravitating disc was earlier studied in detail by Jog (1999) to explain the radius beyond which lopsidedness is seen in spiral galaxies. In that problem, the disc response to a distorted halo was calculated. We can clearly see that
{\it the negative disc response is a general phenomenon valid for any self-gravitating disc subjected to an external perturbation.} 

Note another striking feature that the vertical disc response studied here is independent of the central density $\rho_{d0}$ of the galaxy for a 'disc alone' system and is weakly dependent on the central 
density $\rho_{c}=\rho_{d0} +\rho_{h0}$ for a 'disc plus dark matter halo' system.
This becomes clear from a close examination of eqs.([26],[27]) and eq.(22b) as 
$h(R)$ is inversely proportional to $\rho_{d0}$ in the first case,
and it depends on ${\rho_{h0}/{\rho_{d0}}}$ (which is a small number in general) 
in the second case. So the behaviour of the disc linear response along the 
galactocentric distance depends mainly on the three dimensional 
mass-distribution and depends only weakly on the central brightness of the galaxy. Thus the results obtained below are general and are valid for both normal as well as the low surface brightness galaxies.

\subsection{Self-consistent treatment:}

\noindent In obtaining the disc response potential (see sec.2.1.2) we have ignored 
the self-gravity of the perturbation itself i.e. the disc response potential 
was calculated due to the imposed potential alone. Obviously the stars in the 
disc will be affected by both the externally imposed potential and the 
potential arising due to the disc response to it. So from the requirement 
of self-consistency the total potential ($\Psi_{t}$) that would be 
experienced by the disc can be written, following Jog (1999), as:

$$ \Psi_{t}\:=\: \Psi_{1} + {\Psi^{\prime}}_{resp} \eqno(28)$$

\noindent where ${\Psi^{\prime}}_{resp}$ is the self-gravitating potential 
corresponding to the net self-consistent change in the disc
density ($\rho^{\prime}$) which is obtained as a response to the total 
potential $\Psi_{t}$. Following the argument given by Jog (1999), we can 
write the total potential experienced by the disc in the presence 
of an external, linear perturbation as

$$  \Psi_{t} \:=\: \chi_{m} \Psi_{1} \eqno(29) $$

\noindent where $$\chi_{m} \:=\: \frac{1}{1 + |{\mathcal R}_{m}(R)|} \eqno(30) $$

$\chi_{m}(\leq 1)$ is called the reduction factor or the susceptibility of the disc in a zero-forcing frequency field. It tells us how the magnitude of the perturbing potential is reduced due to the self-consistent negative disc response.
Since $|{\mathcal R}_{m}(R)|$ is a positive definite quantity, the magnitude of the total perturbed potential in the disc plane is always less than that of the external perturbing potential. In the limiting case of
 $|{\mathcal R}_{m}(R)|=0$ implying  $\chi_{m}=1$ corresponds to no reduction effect i.e. the disc self gravity plays no role and hence the disc can be taken to be directly exposed to the externally imposed potential.

\section {Results:}

\noindent We study the radial behaviour of different Fourier components ($m$) of 
the response potential and investigate its implications for astronomical 
observations. The basic idea is to find whether there is a pronounced minimum in the response potential and to see if the position of this minimum which we call $R_{min}$ occurs within the typical size of the galatic disc. The net vertical distortion will be seen beyond this radius as argued next.

The net perturbed motion is set by the total potential $\Psi_t$ (eq.[29])
 and will decide where a warp is seen. This net perturbed
motion is described by eqs.(8) and (9) except that here the 
perturbation potential $\Psi_1$ is replaced by the total potential 
$\Psi_t = \chi_m \: \Psi_1$ which takes account of the self-consistent 
response of the disc.
Note that the term $1/[{\nu}_{\circ}^2 - \Omega_{\circ}^2]$ in equation (9)
increases monotonically with radius for any realistic disc
galaxy, while $\chi_m$ has a minimum at $R_{min}$. Hence the
product of these two terms has a minimum close to $R_{min}$.
Thus the net perturbed motion and hence warp (for $m=1$) and
scalloping (for $m=10$) will only be seen beyond $R_{min}$ for
$m=1$ and $10$ respectively.
Therefore, although the response potential and the reduction factor rise on both sides of the minimum (Figs. 1 and 2), the warps will be seen only in the outer parts. This is
particularly true in a realistic tidal encounter where the
magnitude of the  perturbation
potential $\Psi_{po}$ will be higher at larger radii whereas
here for simplicity we have assumed this to be nearly constant across the disc, see the discussion following eq.(22b).

This has observational relevance if $R_{min}$ occurs within the typical size of a galactic disc as mentioned above.
While the optical or stellar disc in a galaxy was long believed to have a sharp outer truncation at 4.5-5 exponential disc scalelengths (van der Kruit \& Searle 1981), it has been argued recently on general grounds that the disc can extend beyond this (Narayan \& Jog 2003), thus the size of the disc can be larger. In fact, the stellar disc in some galaxies such as NGC 2403 and M33 (Davidge 2003), NGC 300 (Bland-Hawthorn et al. 2005) and M 31 (Ibata et al. 2005), is observed to extend to even 8 or larger disc scalelengths. 
Thus we are justified in analysing the vertical perturbations to stellar discs upto these radii (Section 3.1).

\subsection {\bf $m=1$ component}

The perturbing potential with an $m=1$ component (as in eq.[4b]),
 generates 
the well-known integral-sign warp in a galactic disc. 
 However, so far the physical reason as to where this warp starts
in the disc is not understood (Section 1). 
We try to answer below this question by studying the disc response to 
the $m=1$ component of the tidal perturbation.    

\subsubsection {\bf Radius for onset of warps}

\noindent The dimensionless response potential (eq.[26]) depends mainly on the ratio 
of the scale-height to the scale-length $z_{\circ}/R_{d}$ - which we
define to be  the {\it thickness parameter} of the disc. This is an important parameter because it provides the 
information regarding the three-dimensional structure of the disc. 
A higher value of the thickness parameter indicates a thicker disc, and
conversely it is smaller for a thinner disc. It has a weak dependence on the central density of 
the galaxy in general and in particular for a 'disc-alone' system it is 
independent of the central density (see Section 2.1). For the sake of simplicity
 we first treat a constant value of the thickness parameter across the disc.  
 
In Fig.(1), 
$\mathcal{ R}_{1}(R)$, the response function for the $m=1$ component is 
plotted as a function of radius for a constant value of 
$z_{\circ}/R_d = 0.2$ for the disc-alone case and for a disc
plus an oblate dark matter halo.
 The minimum of this response lies  within the extent of the optical 
exponential disc:  $R_{min}=5.6  R_{d}$ 
 for a disc-alone system, while $R_{min}=5.1 R_{d}$
for the disc plus an oblate-halo system, 
with halo parameters $R_{c}=1.7 R_{d}$, $\rho_{h0}= 0.025$ 
(in units of $M_d/R_{d}^3$) and $q=0.95$. Thus the inclusion of a rigid halo has a marginal effect on the resulting value of $R_{min}$, the reason for this is discussed later in this section. 
 Note that these halo parameters are chosen so that they match 
 closely the halo mass model of our Galaxy (see Mera et.al. 1998) 
 and  the halo mass, $M_h$, is $\sim 7 M_{d}$, the disc mass, for these 
parameters 
 within the total extent of the disc studied ($\leq 10 R_d$). 

In Fig. 2  we plot the reduction factor
$\chi_{1}$ versus radius, which also shows a minimum close to $R_{min}$.
As discussed above, beyond this radius, the magnitude of the
negative disc response decreases, and hence the
net disc response increases.    Thus the self-gravity of the
disc resists a vertical distortion of type $m=1$ in the inner regions and a
net warp is only seen beyond the above $R_{min}$, which is $\sim 5.1
R_{d}$ for the disc plus oblate-halo case. We note that
this lies in the range where warps are seen in galaxies (Briggs 1990).  

\medskip

\noindent {\bf Varitation with thickness parameter, $z_0/R_d$:}

\medskip

\noindent We have scanned a possible range of the ratio $z_{\circ}/R_d$ 
from $0.05$ to $0.5$ in case of disc plus an oblate-halo
 for $M_h \sim 7 M_d$ and in all these cases $R_{min}$ lies within 
$4 R_d$ to $6 R_d$, see Table 1 below for details. {\it Thus we show that the  warps will be seen  beyond a radius in the range of 4-6 $R_d$,
which agrees well with the observed radial range for the onset of warps (Briggs 1990)}.
This shows that in a normal galaxy disc if warps are generated solely 
due to the tidal perturbation they are likely to start between $4-6 R_{d}$.

Further note that that as the thickness parameter, $z_{\circ}/R_d$, increases, the resulting $R_{min}$ 
 for the response lies farther in, in the inner region of the disc.
 The reason for this is that the self-gravity of the disc reduces as the 
 disc becomes puffed up in the vertical direction, hence the net
disc distortion can be seen from an earlier radius. This brings out 
clearly why thin discs have a better resistive 
power against an external perturbation than the thicker discs. 

This physical point has the following implication: if a disc is
already heated say due to internal sources such as clouds or
spiral arms as in late-type galaxies, then any perturbation of the type considered here (namely the $m=1$ case) is
more likely to disturb it into a warp. Thus a thicker disc is more
likely  to show warping. 
This could also be the reason why a larger fraction of galaxies at high redshift are observed to be warped (Reshetnikov et.al. 2002), both because tidal interactions are more likely at high redshift, and these can also heat up a disc, and also because galaxies at high redshift are more gas-rich which can
heat up the discs internally.

\begin{table*}
\caption{$R_{min}/ R_d$ for $m=1$ mode vs. Thickness parameter ($z_{\circ}/R_d$)}
\begin{tabular}{|c||c||c||c||c||c||c|}
\hline 
$(z_{\circ}/R_d)$ & 0.05 & 0.1 & 0.2 & 0.3 & 0.4 & 0.5\\
\hline
$R_{min}/R_{d}$ & 5.9 & 5.6 & 5.1 & 4.7 & 4.4 & 4.2\\
\hline
\end{tabular}
\end{table*}

\subsubsection {\bf Implications and Discussion:}

\medskip

\noindent {\bf Weak Dependence on the halo mass:}

\noindent Figs. 1 and 2 show that the minima in these and hence the radius for the onset of warps is only marginally affected ($\sim$ 10\%) on inclusion of a rigid dark matter halo. This may seem surprising since the halo is known to be dominant in the outer parts. However, there are two reasons for this: first, the disc mass and the halo mass are still comparable to within a factor of few upto the region of interest, namely around $R_{min}$. Second and more important reason is that the halo density distribution is an even function of $z$ and hence on including the halo, the only change in the disc response function occurs due to the 
effect on $H(R)$ in eq.[22] (see Section 2.1.2 for details). This weak dependence on the halo mass is highlighted in the next figure.

In Figure 3 we plot the location of the minimum of the response potential, $R_{min}$, for $m=1$ vs. $M_h/M_d$, the ratio of dark matter halo-to-disc mass (where the ratio is obtained for the mass within the size of the disc considered, namely 10 $R_d$). The most striking result from this figure is that $R_{min}$ has a weak dependence on the halo-to-disc mass ratio.  The value of  $R_{min}$ decreases as the halo becomes more and more massive compared to the disc mass. This is because in the presence of a massive halo and hence a higher restroring force, the outer parts of the disc rises gradually as compared to the case of disc alone system where the disc edge bends sharply. We find that $R_{min}$ for the onset of warps is smaller when the disc edge bends more gradually (Fig. 1). This particular result explains well
the observational point made by Sanchez-Saavedra et al. (2003) that
the farther away the warp stars, the steeper it rises. 

So far we have considered the halo to be rigid. An inclusion of a live halo as in a realistic galaxy is much harder to treat without N-body simulations. 
A live halo would respond to both the direct tidal potential and the 
perturbed disc configuration. The halo response to the tidal field is shown to
magnify the tidal field as seen by the disc and this in turn can increase the amplitude of warp(m=1)(see Weinberg 1995). 
So if the live halo forces the disc edge to bend more sharply then we would expect to see an outward-shift in $R_{min}$. However, in view of the above discussion on $R_{min}$ vs halo mass, we suspect that the location of $R_{min}$ will not be strongly affected in the presence of a live dark matter halo in our problem.   

\medskip

\noindent {\bf Effect of increasing scale-height, and planar random motion on $R_{min}$:}

\noindent The above calculation assumes a constant disc thickness with radius.
In a realistic galaxy, the stellar scale-height flares moderately
within the optical radius (de Grijs \& Peletier 1997, Narayan \&
Jog 2002).
For  such a disc where the  
scale-height or the thickness parameter increases with radius,
we find that the resulting  values of $R_{min}$ lie slightly within $4$ 
disc scale-lengths.
Inclusion of significant planar random motions (which is important for the
stellar disc component) in the disc can also shift the $R_{min}$ values in the inward direction. This can be understood from the fact that the planar random motions make the disc stiffer and warp the disc through smaller angles, that is the disc edge bends more gradually, as was shown by Debattista \& Sellwood (1999).

These can together explain why many discs show stellar warps (e.g. Reshetnikov \& Combes 1998), which therefore must start within the optical radius.

\medskip

\noindent {\bf Comparison with Radius of onset for planar lopsidedness:}

\noindent A thin disc has a stronger self-gravitational force in the
vertical direction than in the plane, which explains why the radial distance
of 4-6 $R_d$ upto which
a disc is able to resist the vertical distortion for the $m=1$ mode as we
have found here, is
larger than the disc resistance upto $\sim 2 R_d$ to the $m=1$ mode in the 
plane (Jog 2000).
Thus the idea of negative disc response explains naturally why
discs are susceptible to onset of planar lopsidedness from  smaller
radii as observed (Rix \& Zaritsky 1995, Bournaud et al. 2005) while they are able to resist warps till a larger radius (Briggs 1990).

\medskip

\subsection {\bf Higher $m$ components}     

Perturbations with higher-order azimuthal wavenumbers ($m > 1$) are of 
particular interest because they may be responsible for producing the
corrugations generally seen in the outer region of disc galaxies
(Florido et al. 1991). For example for the $m = 4$ component, we find 
that the minimum of the response potential is at $R_{min}= 6.5 R_d$ .

The HI gas in the outermost parts of Milky Way shows a remarkable scalloping 
with a high azimuthal wavenumber $m \sim 10$ (Kulkarni et
al. 1982). This interesting feature of the Galactic disc provided us the
motivation for studying the response to an $m=10$ component of the perturbation.
 The gas surface density is larger than the stellar surface density by
about 18 kpc or six disc scalelengths (Narayan et al. 2005),
hence the dominant mass component of the disc is HI gas. The treatment for a
self-consistent disc response derived in this paper is therefore directly
applicable to the gas disc at large radii.
 For the $m=10$ case, the minimum of the response potential, $R_{min}$, 
 occurs at $8.1 R_{d}$ for a disc-alone case, and at $7.1 R_{d}$ for the case of a disc plus an oblate dark matter halo as shown in Fig.4. The dark matter halo parameters used are the same as for Fig.(1). 
 
 It is interesting to see that the strength of the dimensionless disc response
 potential at $R_{min}$ in this case is about $10$ times less than that of 
 the $m=1$ component. But despite its small magnitude, the signature of this higher order perturbation 
 can be clearly seen through the corrugations produced in the midplane 
 density in the outermost part of the disc. In Fig.5 we have
shown the resulting isodensity contours of the midplane density perturbed by 
the $m=10$ Fourier component of the external perturbation. The contours show a clear deviation from the exponential disc embedded in a screened isothermal dark matter halo beyond $7$ disc scale lengths, and the deviation is strong by 10 disc scale-lengths or 30 kpc ( assuming $R_{d} \sim 3$ kpc ). This behaviour  is in a good agreement with the scalloping with $m=10$ observed in the outermost regions of HI distribution in the Galaxy (Kulkarni et al. 1982).

\section{Conclusions}

This paper shows that the self gravity of the disc resists any
vertical distortion in a galactic disc embedded in a dark matter halo.
The main results obtained are as follows:

\noindent 1. We calculate the self-gravitational potential corresponding to  a non-axisymmetric, self-consistent density response of the disc, and show 
that it opposes the perturbation,  and this resistence is
stronger in the inner parts of a galactic disc.
 Thus, the net vertical distortion in the disc is only seen in the outer regions. For example, for the $m=1$ azimuthal wavenumber, the disc
response opposes the imposed perturbation upto $\sim$ 4-6 disc 
scale-lengths, so that the disc shows a net warp only beyond this region.
This physically explains the well-known  observation 
that the onset of warps is typically observed beyond this range of radii (Briggs 1990).
   
\noindent 2. The net effect of vertical
perturbations with higher azimuthal wavenumbers is seen at
higher radii. For example, for $m=10$, the net effect on the disc
can only be seen beyond 7 exponential disc scale-lengths and the effect is 
prominent by 10 disc scalelengths or 30 kpc. The resulting corrugations of the 
mid-plane density explains the long-known puzzle of scalloping detected in HI at these radii in our Galaxy by Kulkarni et al. (1982).

\noindent 3. The disc self-gravitational force is stronger along the vertical
direction than in the plane. Hence the idea of negative disc repsonse
due to the disc self-gravity studied here, and earlier for the planar case (Jog 1999, 2000), naturally explains why discs can resist vertical distortions till
a larger radius than the planar ones. This explains why galaxies show net 
 planar lopsidedness from smaller radii (Rix \& Zaritsky 1995) compared
to the larger radii beyond which warps are seen (Briggs 1990).

\noindent 4. This paper, plus the  earlier work for the
planar disc response (Jog 1999, Jog 2000), shows that the inner regions
of discs are robust and able to resist a distortion as say induced by
tidal interactions. This is a general result and is valid for any disc or a disc plus halo system subjected to an external perturbing potential. This result could be important in the gradual build-up of galactic discs now being studied in cosmological scenarios.

\bigskip

\noindent {\bf Acknowledgements:} We are grateful to the anonymous referee for  very useful and constructive comments, particularly regarding the inclusion of a dark matter halo. K.S. would like to thank the CSIR-UGC, India for a Senior research fellowship.

\newpage

\centerline  {\bf  {References}}

\bigskip

\noindent  Arfken, G. 1985, Mathematical Methods for Physicists ( New York: Academic )

\noindent  Binney, J. \& Tremaine, S. 1987, Galactic Dynamics (Princeton: Princeton Univ. Press) 

\noindent Bland-Hawthorn, J., Vlajic, M., Freeman, K.C., \& Draine, B.T. 
 2005, ApJ, 629, 239

\noindent Bournaud, F., Combes, F., Jog, C.J., \& Puerari, I. 2005, A \& A, 438, 507

\noindent  Briggs, F. 1990, ApJ, 352, 15

\noindent Cohl, H. S., \& Tohline, J. E. 1999, ApJ, 527, 86

\noindent Davidge, T.J. 2003, AJ, 125, 3046 

\noindent de Grijs, R. , \& Peletier, R.F. 1997, A \& A, 320, L21

\noindent  de Zeeuw, T., \& Pfenniger, D. 1988, MNRAS, 235, 949

\noindent Debattista, V.P., \& Sellwood J. 1999, ApJ, 513, L107

\noindent  Hunter, C., \& Toomre, A. 1969, ApJ, 155, 747

\noindent  Florido, E., Battaner, E., Sanchez-Saavedra, M. L., Prieto, M., \& Mediavilla, E. 1991, MNRAS, 251, 193 

\noindent Ibata, R., Chapman, S., Ferguson, A.M.N., Lewis, G., Irwin, M., \& Tanvir, N. 2005, ApJ, 634, 287 

\noindent  Jackson, J. D. 1975, Classical Electrodynamics (New York: Wiley )

\noindent  Jog, C. J. 1999, ApJ, 522, 661

\noindent  Jog, C. J. 2000, ApJ, 542, 216 

\noindent  Kuijken, K., \& Gilmore, G. 1989, MNRAS, 239, 571

\noindent  Kulkarni, S. R., Blitz, L., \& Heiles, C. 1982, ApJ, 259, L63

\noindent  Mera, D., Chabrier, G., \& Schaeffer, R. 1998, A \& A, 330, 953

\noindent  Narayan, C.A., \& Jog, C.J. 2002, A \& A, 390, L35

\noindent Narayan, C.A., \& Jog, C.J. 2003, A \& A, 407, L59

\noindent  Narayan, C.A., Saha, K., \& Jog, C.J. 2005, A \& A, 440, 523 

\noindent  Quiroga, R. J. 1974, Ap \& SS, 27, 323

\noindent  Reshetnikov, V., \& Combes, F. 1998, A \& A, 337, 9

\noindent Reshetnikov, V., Battaner, E., Combes, F., \& Jimenez-Vicente, J. 2002, A \& A, 382, 513

\noindent Rix, H.-W., \& Zaritsky, D. 1995, ApJ, 447, 82

\noindent  Sanchez-Saavedra, M. L., Battaner, E., \& Florido, E. 1990, MNRAS, 246, 458

\noindent  Sanchez-Saavedra, M. L., Battaner, E., Guijarro, A., Lopez-Corredoira, M., \& Castro-Rodriguez, N. 2003, A \& A, 399, 457

\noindent  Sackett, P., \& Sparke, L. 1990, ApJ, 361, 408

\noindent  Schwarz, U. J. 1985, A \& A, 142, 273 

\noindent  Schwarzkopf, U., \& Dettmar, R.-J. 2001, A \& A, 373, 402

\noindent  Terquem, C. 1998, ApJ, 509, 819

\noindent van der Kruit, P., \& Searle, L. 1981, A \& A, 95, 105

\noindent  Weinberg, M. D. 1995, ApJ, 455, L31

\noindent  Zaritsky, D., \& Rix, H. -W. 1997, ApJ, 477, 118

\bigskip

\newpage

\bigskip
\begin{figure}
{\rotatebox{270}{\resizebox{8cm}{8cm}{\includegraphics{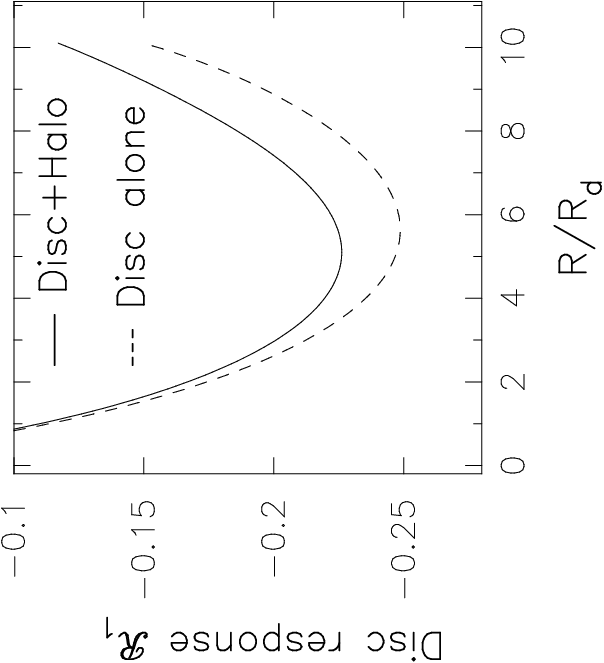}}}}

\noindent {\bf Figure 1.} Dimensionless response potential, $\mathcal{R}_1$ for the $m=1$ Fourier component of the external perturbation for a constant thickness disc. The scale-height to scale-length ratio i.e. $z_{\circ}/R_d$ is taken to be $0.2$. This figure shows that $R_{min}$ occurs at $5.6 R_d$ for the disc-alone system and at $5.1 R_d$ for the disc plus halo system.  
\end{figure}

\begin{figure}
{\rotatebox{270}{\resizebox{8cm}{8cm}{\includegraphics{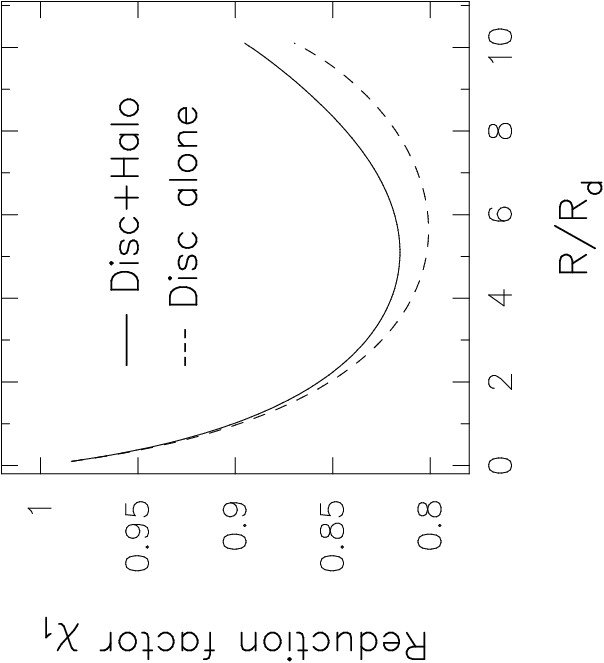}}}}

\noindent{\bf Figure 2.} Reduction factor $\chi_{1}$ for the $m=1$ component of the perturbation potential due to the negative disc response for $z_{\circ}/R_d = 0.2$. The minimum of the reducton factor occurs at a radius $R_{min}$ = $5.6 R_d$ for the disc-alone sysem and at $5.1 R_d$ for the disc plus halo system. Beyond  $R_{min}$, the reduction factor increases steadily.  
\end{figure}

\begin{figure}
{\rotatebox{270}{\resizebox{8cm}{8cm}{\includegraphics{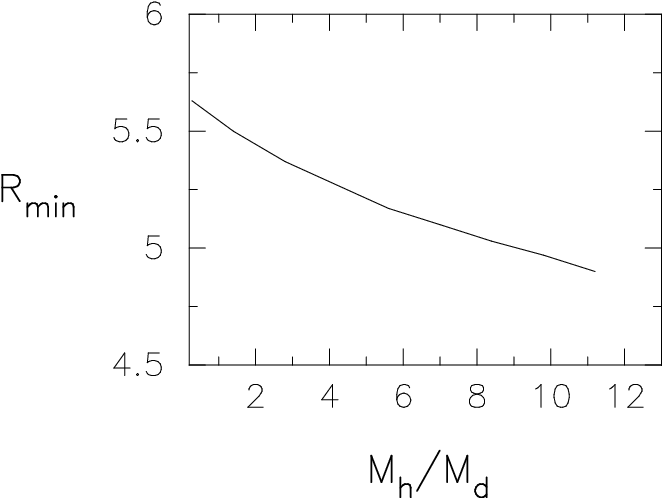}}}}

\noindent{\bf Figure 3.} $R_{min}$ for the reponse potential of $m=1$  vs.
 the ratio of the dark matter halo to disc mass, considered within the 
radius of $10 R_d$. The figure shows that the $R_{min}$ has a weak dependence on$M_h/M_d$, the halo-to-disc mass ratio. 
\end{figure}

\begin{figure}
{\rotatebox{270}{\resizebox{8cm}{8cm}{\includegraphics{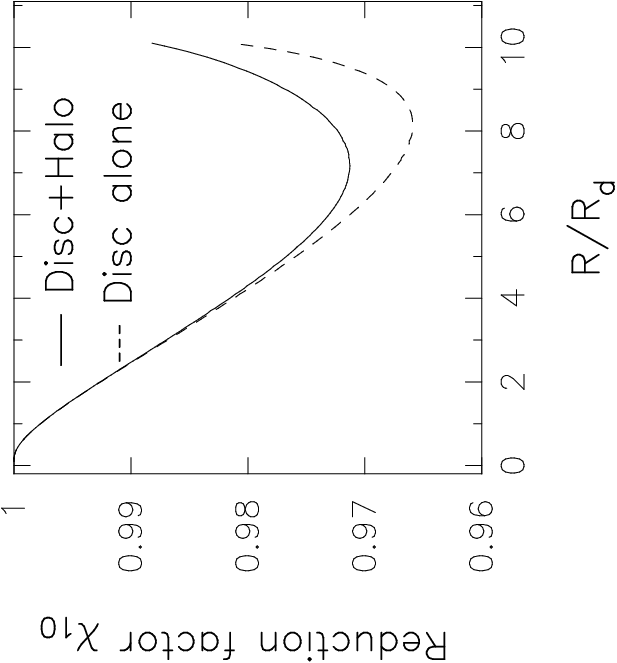}}}}

\noindent{\bf Figure 4.} Reduction factor $\chi_{10}$ for $m=10$ Fourier component of the external perturbation, calculated for $z_{\circ}/R_d = 0.2$. $R_{min}$ of the response is located at $7.1 R_d$ and $8.1 R_d$ for a disc plus halo 
system and a disc-alone system respectively. The reduction factor increases sharply beyond these radii. Note that the reduction due to the negative disc response is very small in magnitude compared to that for the $m=1$ case. 
\end{figure}

\begin{figure}
{\rotatebox{270}{\resizebox{8cm}{8cm}{\includegraphics{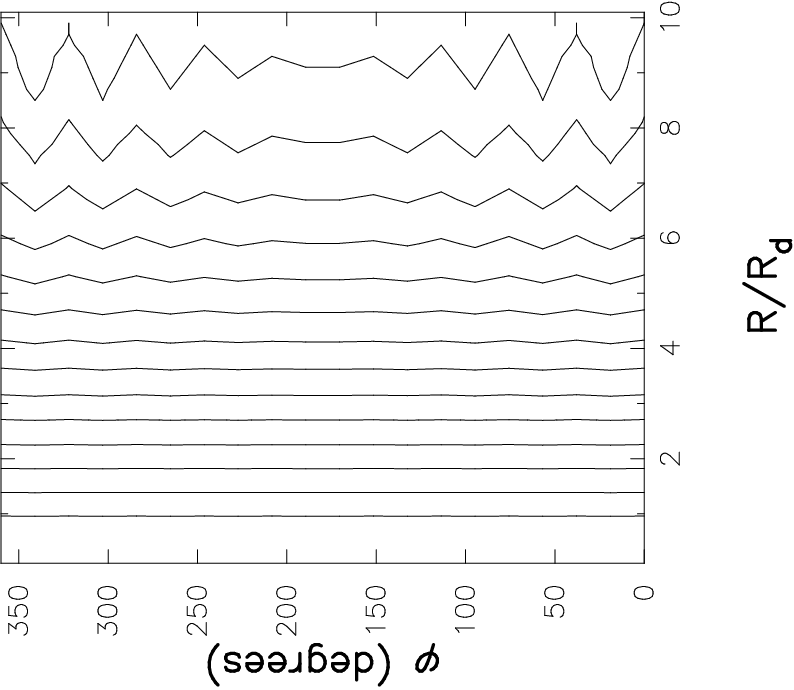}}}}

\noindent{\bf Figure 5.} Contour diagrams for the disc mid-plane density perturbed by the $m=10$ Fourier component of the perturbation in the ($R,\varphi$) plane. The mass of the perturber is taken to be $10\%$ of the disc mass and at a distance $\sim 36$ kpc (for $R_d \sim 3$ kpc) from the disc centre. Contour levels are $\rho_{c}\times(0.408, 0.272, 0.181,0.12,0.08,...)$; where $\rho_{c}$ is the total central mid-plane density of the galaxy in $M_{\odot}pc^{-3}$. Contours after $7$ disc scale-length deviate clearly from the unperturbed midplane mass
 distribution and the deviation is stronger by $\sim$ 10 disc scale-lengths or 30 kpc, showing the signature of scalloping.
\end{figure}

\end{document}